# Numerical Study of CO$_2$ Conversion to SAF in a Fixed Bed Catalytic Reactor

Ruiqin Shan[1], Shengwei Ma[1], Van Bo Nguyen[1], Chang Wei Kang[1], Teck-Bin Arthur Lim[1]

[1]Institute of High Performance Computing (IHPC), Agency for Science Technology and Research (A*STAR), 1 Fusionopolis Way, #16-16 Connexis, Singapore 138632, Republic of Singapore

## ABSTRACT

CO$_2$ hydrogenation to hydrocarbon refers to an indirect pathway of CO$_2$ utilization. Among them, the conversion of CO$_2$ with green H$_2$ to sustainable aviation fuel (SAF) with high energy density has gained much attention. It offers a promising way to reduce greenhouse gas emissions, address the fossil fuel crises, and transform a climate killer into valuable products. However, this low-carbon technology is intrinsically complicated. It involves the development of a catalyst, the design of a reaction system, and its operation and product refining. Hence, it is important to understand the chemical process of CO$_2$ hydrogenation in the reactor. In this study, numerical simulations of a fixed bed catalytic reactor for CO$_2$-to-SAF conversion are conducted by coupling CFD with heterogeneous catalytic reactions at the catalytic surface. The heat and mass transfer between the catalyst surface and surrounding fluid flow are resolved in the simulation. A detailed understanding of the reacting flow and catalytic processes is obtained from this study. The impact of operating parameters, i.e., temperature, pressure, mass flow rate, and the ratio between CO2 and H2, is also explored, which provides important insights into the catalytic reactor design and operation.

## 1. Introduction

Carbon capture and utilization are essential in global efforts to mitigate greenhouse emissions and achieve sustainable development. Carbon capture is a process designed to capture CO$_2$ emissions from sources before they are released into the atmosphere in the energy and industrial sectors [1, 2]. The captured CO$_2$ can then be stored underground in geological formations. Carbon capture alone is financially demanding for industry. Carbon utilization aims to offset the cost by transforming CO$_2$ to value-added products, such as building materials, valuable chemicals, and high energy fuels, which can also create revenue streams and economic opportunities.

Aviation is a significant contributor to greenhouse emissions, and CO$_2$ conversion to sustainable aviation fuel (SAF) has been a very promising research topic for reducing the carbon footprint of aviation and dependence on fossil fuels [3]. For the conversion of CO$_2$ to SAF, it is critical to understand the physical and chemical processes to facilitate the chemical system design, catalyst design, and performance optimization. However, the related research is very limited due to the intrinsic complexities for multi-scale and multi-physics problems. In the present paper, numerical simulations are conducted for a fixed-bed reactor to resolve the catalytically reacting flow and interactions between the flow and catalytic pellets. The simulations are expected to provide a detailed understanding of the chemical processes of CO$_2$-to-SAF conversion and the effects of operating conditions.

## 2. Numerical Methodologies
### 2.1 Governing Equations and Numerical Methods

A CFD tool for simulating the CO$_2$-to-SAF process is developed here based on the solver catalyticFOAM [4]. The solver allows for CFD simulations of reactive flow involving both gas-phase reactions and heterogeneous reactions at catalytic surfaces. For the gas-phase bulk flow, the governing equations that describe the conservation of total mass, momentum, species, and energy are shown below:

$$\frac{\partial \rho}{\partial t} + \nabla \cdot [\mathbf{u}\rho] = 0, \quad (1)$$

$$\frac{\partial (\rho \mathbf{u})}{\partial t} + \nabla \cdot [\mathbf{u}(\rho \mathbf{u})] + \nabla p + \nabla \cdot \mathbf{T} = \mathbf{0}, \quad (2)$$

$$\frac{\partial (\rho Y_i)}{\partial t} + \nabla \cdot [\mathbf{u}(\rho Y_i)] + \nabla \cdot \left[ -D_{i,m} \nabla (\rho Y_i) - D_{T,i} \frac{\nabla T}{T} \right] = \dot{\omega}_i^{hom}, i = 1, \dots, N-1, \quad (3)$$

$$\rho \overline{C_p} \frac{\partial T}{\partial t} + \rho \overline{C_p} \mathbf{u} \nabla T + \nabla \cdot [\mathbf{u}p] + \nabla \cdot (\mathbf{T} \cdot \mathbf{u}) + \nabla \dot{q} = \dot{\omega}_T^{hom}. \quad (4)$$

In the above equations, $\mathbf{T}$ is the viscous stress tensor. $Y_i$ and $D_{i,m}$ are the mass fraction of the $i$-th species and the mass diffusion coefficient for the $i$-th species in the mixture, respectively. $D_{T,i}$ is the thermal (Soret) diffusion coefficient of the $i$-th species. $N$ represents the number of gas-phase species in the reaction mechanism. $\dot{\omega}_i$ is the mass production rate of the $i$-th species. In the energy equation, $\overline{C_p}$ is the mixture-averaged specific heat capacity at constant and $\dot{q}$ is the heat flux from heat conduction and mass diffusion of species with different enthalpy. $\dot{\omega}_T$ is the chemical heat release rate due to the gas-phase reactions. The equation of state $p = \rho R T$ is used here for the calculation of pressure, where $R$ is the specific gas constant of the mixture.

For heterogeneous surface reactions, the governing equation that describes the conservation of the site fraction of adsorbed surface species is shown below:

$$\sigma_{cat} \frac{\partial \theta_i}{\partial t} = \omega_i^{het}, i = 1, \dots, M-1, \quad (5)$$

where $\theta_i$ is the site fraction of the $i$-th adsorbed species and $\omega_i^{het}$ is the production rate of the species from the heterogeneous reactions. $\sigma_{cat}$ represents the site density of the catalyst, and $M$ is the number of adsorbed species in the heterogeneous reaction mechanism. The interaction between the gas-phase bulk flow and the catalyst takes place on the catalytic surface. The mass flux and heat flux at the catalytic surfaces are calculated based on the heterogeneous reactions.

The governing equations are solved using segregated algorithms with an operating-splitting method, which applies to species equation and energy equations involving chemical reactions. The reaction operator is separated from the transport operator allowing each

---

Corresponding author: Ruiqin Shan
*E-mail address*: shanr@ihpc.a-star.edu.sg

operator to employ the most suitable numerical method. Details of the methodologies used in the solver can be found in the reference [4].

## 2.2 Kinetic Model

SAF, as an alternative jet fuel, has almost identical physical and chemical properties to conventional jet fuel based on fossil fuels while featuring low-carbon emissions. SAF can reduce $CO_2$ emissions by up to 80% compared with fossil fuels [5]. In terms of compositions, jet fuel consists of hundreds of hydrocarbons with carbon number ranging from 8 to 16, which are mainly paraffins [6]. The accurate composition varies based on feedstock sources and refining processes. The pathway to SAF production from $CO_2$ generally consists of two major steps, i.e., the reverse water gas shift reaction (RWGS) and the Fisher-Tropsch (FT) synthesis [6]. In the present paper, the reaction model for the hydrogenation of $CO_2$ to hydrocarbons is obtained from the reference [7], as shown below:

$$CO_2 + H_2 \leftrightarrow CO + H_2O \qquad (6)$$
$$CO + 2H_2 \leftrightarrow (CH_2) + H_2O \qquad (7)$$

where $(CH_2)$ represents long chain hydrocarbons here. The kinetic model and parameters shown in Eqs. (8-11) and Table 1 are obtained through experiments in a fixed-read reactor loaded with an alumina supported iron catalyst Fe/K@γ-$Al_2O_3$.

$$r_{RWGS} = \frac{k_{RWGS}\left(P_{CO_2}P_{H_2}^{0.5} - \frac{P_{CO}P_{H_2O}}{K_{eq}P_{H_2}^{0.5}}\right)}{(1 + a_{RWGS}P_{H_2O}/P_{H_2})} \qquad (8)$$

$$r_{FT} = \frac{k_{FT}P_{H_2}P_{CO}}{\left(1 + \frac{a_{FT}P_{H_2O}}{P_{H_2}} + b_{FT}P_{CO}\right)} \qquad (9)$$

$$k_i = k_{i,ref}\exp\left(-\frac{E_{A,i}}{R}\left(\frac{1}{T} - \frac{1}{T_{ref}}\right)\right) \qquad (10)$$

$$\log K_{eq} = 3.933 - \frac{4076}{T - 39.64} \qquad (11)$$

where $K_{eq}$ is the equilibrium constant and T is in unit Kelvin. $p$ is the partial pressure of different species. Details about the development of the kinetic model and the catalyst characteristics can be found in the reference [7].

Table 1 Kinetic parameters for the kinetic model in Eqs. (8-11) [7].

| Parameter | Unit | Estimated value |
| --- | --- | --- |
| $k_{RWGS,300\,°C}$ | mol h$^{-1}$ g$^{-1}$ bar$^{-1.5}$ | $8.13 \times 10^{-2}$ |
| $k_{FT,300\,°C}$ | mol h$^{-1}$ g$^{-1}$ bar$^{-2}$ | $6.39 \times 10^{-2}$ |
| $E_{A,RWGS}$ | kJ mol$^{-1}$ | $1.15 \times 10^{2}$ |
| $E_{A,FT}$ | kJ mol$^{-1}$ | $6.78 \times 10^{1}$ |
| $a_{RWGS}$ | – | $1.63 \times 10^{1}$ |
| $a_{FT}$ | – | 9.07 |
| $b_{FT}$ | bar$^{-1}$ | 2.44 |

## 2.3 Physical Model

The reactor model used in the present paper is a typical fixed-bed tubular reactor loaded with catalyst pellets. It is simplified to a 2-dimensional rectangular channel as shown in Fig. 1 to save computational cost while still being adequate to study the physical and chemical processes inside the reactor. The reactor model is 50 cm long and 2.54 cm wide. The catalyst pellets have a diameter of 5 mm.

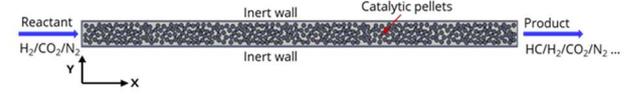

Fig. 1 Computational configuration for the fixed-bed reactor loaded with catalyst pellets.

The reactant enters the reactor through the inlet and the product exits through the outlet. The top and bottom walls are inert walls with a fixed temperature, while the surfaces of the catalyst pellets are catalytic walls where heterogeneous reactions take place. The interior of the catalyst is not resolved here, and the gas mixture flows through the void space in the reactor.

## 3. Results and Discussion

In the present study, the reactor is fed with $CO_2$, $H_2$, and $N_2$. The operating conditions include the molar ratio between $CO_2$ and $H_2$, pressure and temperature of the reactant, and inlet velocity. A parametric study is conducted on the operating conditions. The conditions for the baseline case are: $(H_2/CO_2)_{mol}$ = 9, pressure = 20 bar, temperature = 900 K, and inlet velocity = 0.1 m/s. The temperature of the top and bottom walls is fixed and set to be the same as the feed gas temperature, i.e., 900 K for the baseline case. The flow field of the baseline case and the sensitivity of the chemical process to the operating conditions will be discussed in this section.

### 3.1 Grid Independence Study

A grid independence study is conducted under the operating condition for the baseline case. Four mesh grid sizes are tested with the total number of cells being 56000, 80000, 188000, and 372000, respectively. Comparisons of the averaged properties, including temperature and key chemical species, along the reactor length with different grid sizes are plotted in Fig. 2. The profiles are obtained by averaging the flow properties over the cross-sectional area of the reactor when the simulation reaches a steady state. As shown in Fig.2, there are only negligible differences in these important properties from coarser mesh to finer mesh. The grid size with 80000 cells is selected to ensure both computational accuracy and efficiency.

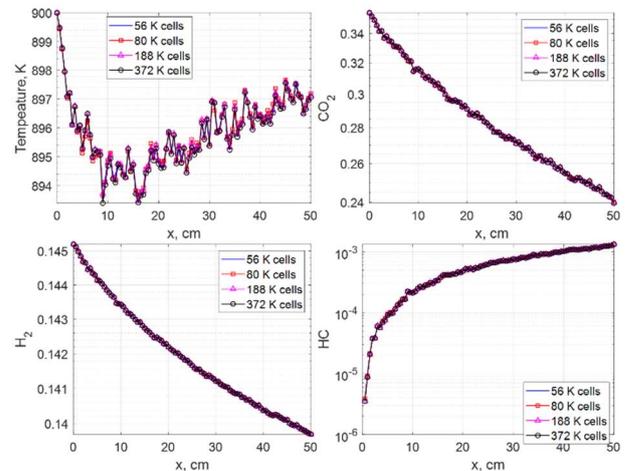

Fig. 2 The comparisons of averaged temperature, mass

fraction of $CO_2$, $H_2$ and hydrocarbon (HC) profile along the reactor with different grid sizes of meshing.

### 3.2 Flow Field of the Baseline Case

The reactor is initially filled with a non-reactive mixture of $H_2$ and $N_2$ at a pressure of 20 bar and a temperature of 900 K. The feed gas contains 50% $N_2$ by mass. The time evolution of the hydrocarbon mass fraction is shown in Fig. 3 for the baseline case. As the mixture of $CO_2$, $H_2$, and $N_2$ enters the reactor from the inlet, the two reactions, RWGS and FT synthesis, take place on the surface of the catalyst pellets. $CO_2$ and $H_2$ reacts to form CO which subsequently reacts with $H_2$ to produce hydrocarbons. As shown in Fig.3, the mass fraction of hydrocarbons increases from the inlet to the outlet and with time as well until the flow reaches a steady state.

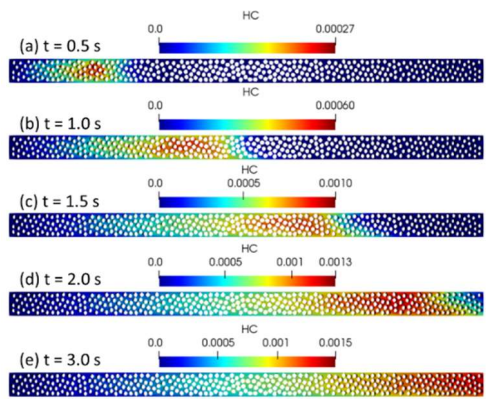

Fig.3 Time evolution of flow field of hydrocarbon mass fraction for the baseline case.

Figure 4 depicts the profiles of the area-averaged mass fraction of chemical species with time at the outlet. At the initial stage, the amount of $H_2$ remains the same because the reactive gas has not yet reached the outlet, and the reactor is initially filled with a non-reactive mixture of $H_2$ and $N_2$. The mass fraction of $H_2$ starts to drop, and the other species increase rapidly at t = 2 s when the reactive feed gas and upstream products arrive at the outlet. The average properties achieve a steady state at around t = 3 s.

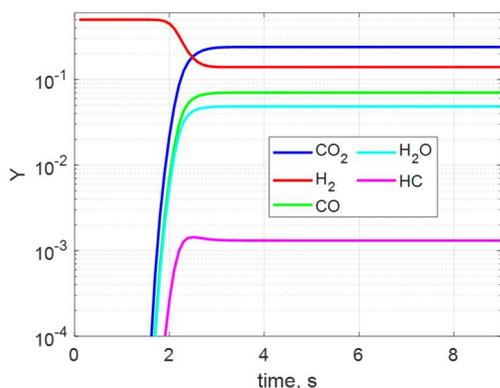

Fig.4 Time evolution of area-averaged mass fraction of chemical species at the outlet for the baseline case.

The steady-state flow fields of the mass fraction of chemical species and the temperature are shown in Fig. 5. The consumption of $CO_2$ and $H_2$ and the production of CO, $H_2O$, and hydrocarbons can be clearly observed from the contours. The more catalytic surface area the feed gas passes through from the inlet to the outlet, the more heterogeneous reactions take place on the surface. For the temperature, it can be seen in Fig. 5 that it drops up to 10.2 K in the interior of the reactor due to endothermic RWGS reaction. Although the FT synthesis reaction is exothermic, the heat absorbed by the RWGS reaction exceeds the heat release under the present operating conditions and catalyst loading. The temperature near the top and bottom walls is higher than in the interior due to the fixed temperature boundary condition set for the walls.

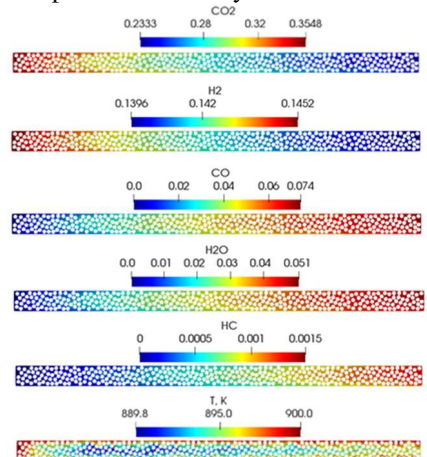

Fig.5 Flow field of temperature and species mass fractions at steady state for the baseline case.

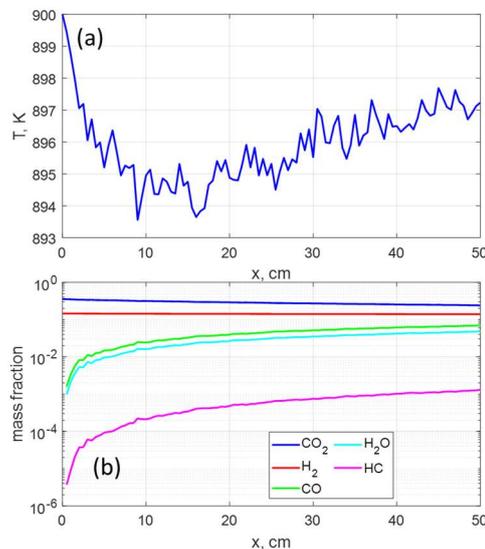

Fig.6 The profile of area-averaged (a) temperature and (b) mass fraction of chemical species at steady state along the reactor length for the baseline case.

The steady-state profiles of the averaged temperature and chemical species mass fraction are given in Fig. 6. It can be seen that the temperature starts to drop rapidly at the inlet by up to 6.4 K due to the intensive endothermic RWGS reaction near the inlet. The temperature gradually rises as more CO is produced and reacts with $H_2$ to form hydrocarbons, which release heat into the reactor. For the chemical species, it is seen in Fig. 6b that the amounts of

CO$_2$ and H2 decrease along the reactor, while hydrocarbon, CO, and H$_2$O increase due to the hydrogenation process on the catalytic surfaces. It is noted that the decreasing and increasing trend of feed and product gases in Fig. 6b continues to the outlet. In addition, it can be found that there is still a significant amount of CO$_2$ and H$_2$ left at the outlet, indicating that the conversion of CO$_2$ is not yet complete and will continue with longer reactors and more catalyst pellets.

**3.3 Effects of Operating Conditions**

A range of different operating conditions is employed to study their effects on the flow field of the reactor and the conversion performance. The results can provide insight into the optimization of the chemical process. Figure 7 shows the profiles of CO$_2$ conversion and hydrocarbon selectivity at the reactor outlet under different conditions, including pressure, temperature, inlet velocity, and molar ratio between H$_2$ and CO$_2$.

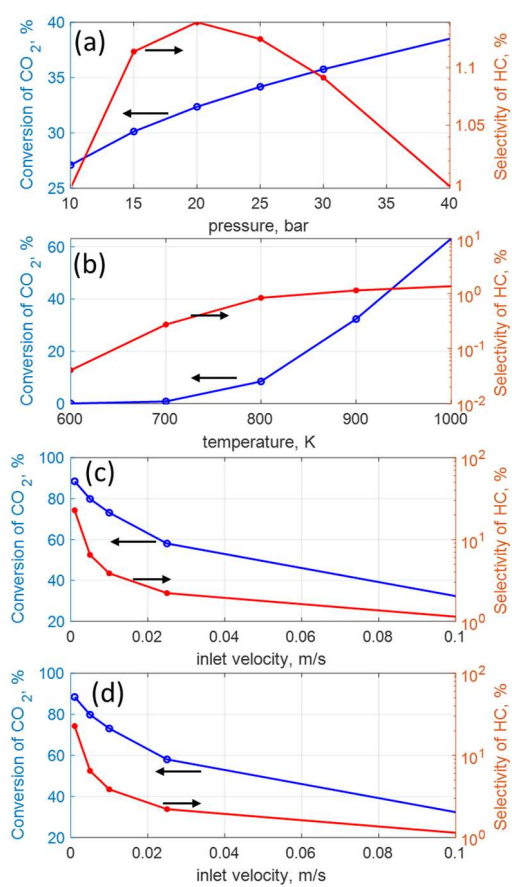

Fig. 7 Conversion of CO$_2$ and selectivity of hydrocarbon at different operating conditions.

In general, the conversion of CO$_2$ and the selectivity of hydrocarbon increase monotonically with higher pressure, higher temperature, lower inlet velocity or longer residence time, and a higher ratio of H$_2$ and CO$_2$ within the present studied range. For pressure effect, the conversion of CO$_2$ reaches a maximum at 20 bar within the studied range of 10 to 40 bar. The behavior of CO$_2$ with respect to pressure is attributed to the kinetic model of RWGS in Eq. (8). The partial pressure of CO$_2$ and CO can affect the reaction direction. With more CO produced, the water gas shift reaction dominates leading to further CO$_2$ reforming.

**4. Concluding Remarks**

The present paper studies the chemical process of CO$_2$ conversion to SAF using CFD simulations. A kinetic model with two-step heterogeneous reactions for CO2 hydrogenation is implemented into the solver for catalytically reactive flow. A rectangular channel loaded with catalytic pellets is employed to mimic a fixed-bed reactor. The simulations provide insights into the hydrogenation process of CO$_2$ in the reactor facilitating a more detailed understanding of the flow field. A series of different operating conditions is applied to the reactor to understand their effects on the flow field and reactor performance, supporting reactor design. It is also worth noting that the simulation results are not compared with experimental data in the present paper, which will be conducted in future studies with available detailed reactor setup and catalyst characteristics, e.g., loading density and surface area.

**5. Acknowledgements**

This work is supported by the RIE2025 USS LCER PHASE 2 PROGRAMME HETFI DIRECTED HYDROGEN PROGRAMME under Grant ID U2305D4001, and Carbon Capture and Utilisation Translational Testbed (CCU-TT) Front-End Loading Phase 3 (FEL 3) under Grant ID C220415017.